\documentclass[english]{article}
\pdfoutput=1
\usepackage[latin9]{inputenc}
\usepackage{url}
\usepackage{graphicx}
\usepackage{amsmath}
\usepackage{amsfonts}
\usepackage{bbm}
\usepackage{color}

\newcommand{\E}{\mathop{\mathbf E}}

\usepackage{babel}

\begin{document}

\title{Optimizing a basket against the efficient market hypothesis}

\author{Fr\'ed\'eric Abergel\footnote{frederic.abergel@ecp.fr}$\,\,$\thanks{BNP Paribas Chair of Quantitative Finance, MAS Laboratory, Ecole Centrale Paris, F-92295 Ch\^atenay-Malabry, France.}$\,\,\,$ and Mauro Politi\footnote{mauro@nt.icu.ac.jp}$\,\,^\dag$\thanks{Department of Economics and Business, International Christian University, 3-10-2 Osawa, Mitaka, Tokyo, 181-8585 Japan.}}

\maketitle
\begin{abstract}
The possibility that the collective dynamics of a set of stocks could lead to a specific basket violating the efficient market hypothesis is investigated. Precisely, we show that it is systematically possible to form a basket with a non-trivial autocorrelation structure when the examined time scales are at the order of tens of seconds. Moreover, we show that this situation is persistent enough to allow some kind of forecasting. 
Market efficiency; Lead-lag correlation; optimization
\end{abstract}

\newpage
\section{Introduction}\label{sec:introduction}
The efficient market hypothesis is technically interpreted to mean that the market is impossible to beat and
that there are no autocorrelation (such as systematically
repeated price/returns patterns) that can be used for
profit \cite{Mandelbrot1966}. The idea is a pillar and a paradigm of classical market finance and it is widely discussed in the influential review of E.~Fama in Ref.~\cite{Fama1970}, dated 1970.  A Markov market, i.e. the assumption that future fluctuations depend only on the last observed price, satisfies the condition of a market that is strictly impossible to beat and thus perfectly efficient. In more recent decades, with advancements in trading technologies, has been clear that the EMH can be broken; a comprehensive survey and discussion can be found in Ref.~\cite{Markiel2003}. However, real markets remains very hard to beat and models that generate no autocorrelation in the price increments are very good first step approximations. In such models, in the limit of small time intervals, the correlations of the increments $x(t_1)-x(t_1-T_1)$ and $x(t_2+T_2) - x(t_2)$ for a generic price series $x(t)$ vanish
\begin{equation}
\left< (x(t_1)-x(t_1-T_1))(x(t_2+T_2) - x(t_2)) \right> = 0\,
\label{eq:emh}
\end{equation}
for each set of finite instants $t_1,T_1,t_2,T_1$ if there is no overlap in the intervals, i.e. if $(t_1-T_1,t_1)\cap (t_2,t_2+T_2) = \emptyset$. This condition is much weaker, and more pregnant, than the 
assumption of stochastically independent increments. Eq.~(\ref{eq:emh}) means that nothing that happened during a time interval can be used to systematically forecast the returns in a future time interval, at least at the level of simple averages and pair correlations. That is, the market is ``effectively
efficient'' in the sense that it is impossible to forecast its direction, paving the way for the application of semi-martingale processes. Contrarily to some other well known stochastic processes such as the fractional Brownian motion, there is no memory in pair correlations to be exploited. In general, this does not rule out higher order correlations or cross-correlations between assets that might be used for technical trading. A Markovian market is ``efficient'' in the strictest sense: it is impossible to beat. Instead, a martingale market leaves the opportunity of exploiting high order dependencies.

It is an acknowledged, tested and recognized fact that individual asset time series do not present features challenging the efficiency - were it not for mean-reverting effects at very short time scales (some seconds), mainly due to the bid-ask bounce. 
The main question we are addressing in this work is whether interdependences between distinct stock price increments can be employed to produce a single time series presenting a non vanishing value of autocorrelation. Basically this is not a new idea - in fact, the ideas beyond pair trading, i.e. co-integration \cite{Johansen1988}, and most of the statistical arbitrage strategies, are simpler versions of this one. The originality of the work we present here lies, firstly, in the focus on short time scales (at the order of tens of seconds) and, secondly and more important, in the design of an optimized collective dynamics. 
Precisely, starting with an original set of $N$ stock time series $x_i(t)$ ($i=1,\ldots,N$), we are asking whether it is possible to find a suitable set of weights $w_i$ for which the basket 
\begin{equation}
B(t) = \sum_{i=1}^N w_i x_i(t)
\label{eq:basket}
\end{equation}
presents systematic autocorrelations, or, more generally speaking, a non trivial persistent behaviour. 

Unfortunately, assessing the persistent or anti-persistent behaviour of a time series raises various difficulties, and the vast confusion in the literature does not help to overcome them. Often, an estimation of the Hurst exponent is recognized as a good indicator, but this is misleading. Strictly speaking, the Hurst exponent is just a measure of the scaling properties of a time series and, alone, does not give any information on the possible persistent behaviour when the analyzed data are not stationary, as is typically the case in the financial world. Moreover, the estimation of the Hurst exponent is by itself a difficult task, and there is a host of available estimators which seldom allow for cross-validations (see, for example, Ref.~\cite{Mielniczuk2007}). 

To avoid these problems, or at least to minimize them, we will mainly focus our discussion directly on the autocorrelation, or, more precisely, on the anti-autocorrelation. This choice will help us overcome the tedious and not always solvable issues of considering trends, leaving us only with the problem of non-stationarity in higher order fluctuations. This could still has an effect on the value of the sample autocorrelation, but we will not focus on that value itself. Rather, we simply seek to assess whether it is significant or not, using tests based on the stationarity of the increments. A consistent way to estimate the statistical significance for non stationary time series is absent, at least to the authors' knowledge.

The paper is organized as follow. In section 2 we present the dataset we use for the study, together with the sampling rules. In section 3 the main statistical objects (autocorrelation and Hurst exponent) are presented and discussed. Section 4 is devoted to the explanation of the tools and procedures we follow looking for the desired non-efficient basket. Finally, section 5 and 6 are dedicated respectively to the results analysis and the conclusions. 

\section{Data set}
\label{sec:data}
\begin{figure}
\centering
\includegraphics[height=5.5cm, width=.95\columnwidth]{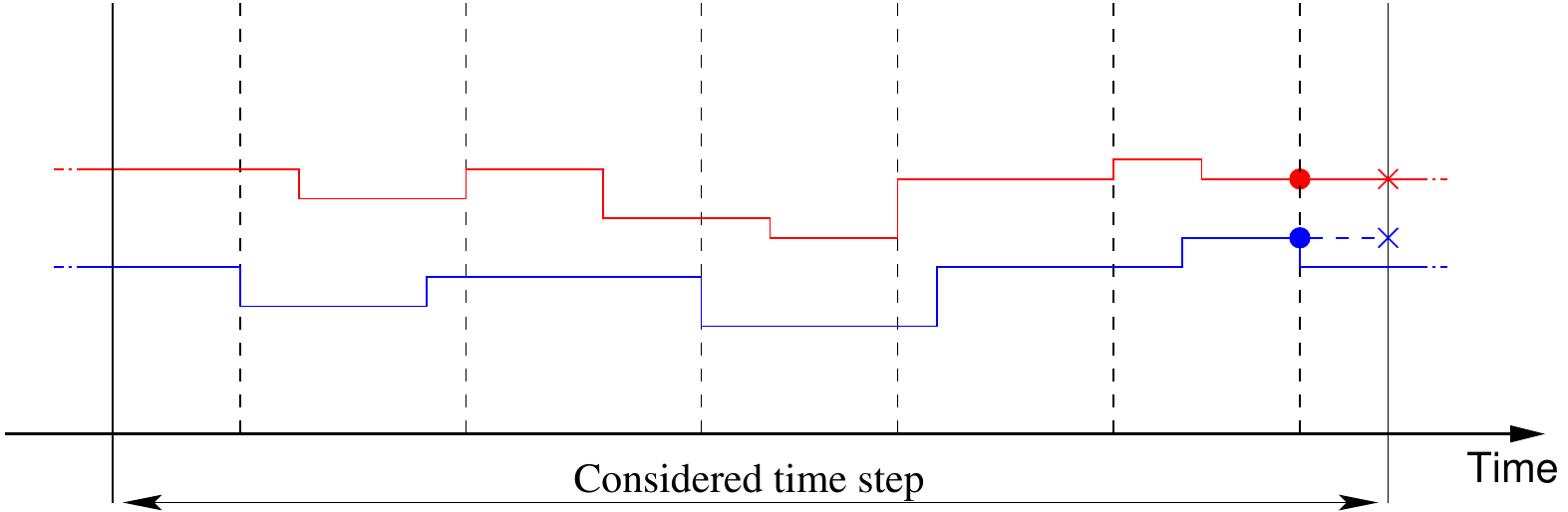}
\caption{Scheme explaining the nature of the series we study and the procedure used to obtain them. For each time window, we observe the last trade (trades are dashed vertical lines, randomly spaced) and consider the bid and ask prices immediately before the transaction (blue and red large points). The blue line is the bid time series and the red one is the ask time series. The considered midprice is simply the average of the considered bid and ask prices.}
\label{fig:midprice}
\end{figure}

The original data set consists of all trades registered in the primary
markets of the analyzed stocks. The data are stored in the Thomson Reuters RDTH data base made available to the Chair of Quantitative Finance by BNP Paribas.
For the purpose of our study, we extract from the RDTH database records consisting in the time of a transaction,
the bid and ask prices prior to each transaction, and the
traded price. These data, appropriately filtered in order to remove
misprints in prices and times of execution, correspond to the trades
registered at NYSE or at NASDAQ during 2007, for the 30 shares of
the Dow Jones Industrial Average Index, namely, at that time: AA.N
AIG.N AXP.N BA.N C.N CAT.N CCE.N DD.N DIS.N GE.N GM.N HD.N HON.N HPQ.N
IBM.N INTC.O JNJ.N JPM.N MCD.N MMM.N MO.N MRK.N MSFT.O PFE.N PG.N
T.N UTX.N VZ.N WMT.N XOM.N. 28 of those companies were traded primarily
at NYSE. IBM.O and MSFT.O are the only two traded primarily at NASDAQ.
The full meaning of the symbols is available from {\tt www.reuters.com}.
The choice of one year of data is a trade-off between the necessity
of managing enough data for significant statistical analyses and the goal of minimizing the effect of strong macro-economic fluctuations. However, the consistency of the discussed results during
extreme condition periods are beyond the purposes of the present paper,
and are left for future studies.

Precisely, for each day the period we consider starts at $10.00am$ and ends at $3.45pm$, leading to $4140$ increments when considering fixed time steps of $5s$, which will be our basic time scale unit. The choice of restricting the considered periods only to the central part of the trading day, discarding the opening and closing period, is justified by the anomalies data often exhibit during these parts of the trading day: errors tend to occur more often during the first and last part of the continuous trading day, it often happens that some shares are open to trading several minutes after the others, due to potential issues during the opening auction. Moreover, as is well known, the activity near the opening and closing period is much higher than during the central part of the trading day, adding a non trivial problem to our study that is strictly based on physical time steps. One could try to work in event time, following a multidimensional approach to trading time that has been recently proposed and successfully used to establish some joint distributional properties of baskets of stocks Ref.~\cite{HuthAbergel2010}, but we thought that the added complexity of this change of time would actually conceal rather than emphasize the idea we are putting forward in this work.

Having fixed a time window, we consider the mid-prices at the time of the last registered trades. Fig.~\ref{fig:midprice} gives a comprehensive explanation of the procedure. We would like to point out to the readers attention the fact we are not working with log-returns, but only with increments.

\section{Autocorrelation}\label{sec:autocorrelation}
Several definitions of the sample autocorrelation coefficients have been proposed in the literature.
We consider the most standard one (see Ref.~\cite{Anderson1941} for a classical discussion and Ref.~\cite{Kan2010} for a modern one): given n observations of a discrete time series $y_1,\ldots,y_n$
the sample autocorrelation coefficient at lag $k$ is given by
\begin{equation}
\hat{\rho}(k) = \frac{\sum_{j=1}^{n-k} (y_i - \bar{y})(y_{i+k} - \bar{y})}{\sum_{i=1}^n (y_i-\bar{y})^2}\,,
\label{eq:autocorrest}
\end{equation}
where $\bar{y} = (\sum_{i=1}^n y_i)/n$ is the sample mean, and $1\leq k \leq n-1$. 
Under the hypothesis of having $y_1,\ldots,y_n$ IID $N(0,\sigma)$ the lag one ($k=1$) autocorrelation can be rewritten for large $n$ as
\begin{equation}
\hat{\rho}(k) = \frac{\sum_{j=1}^{n-1} y_i y_{i+1}}{\sum_{i=1}^n y_i^2}\,.
\end{equation}
The numerator has a null expectation value and a variance 

\begin{eqnarray}
\mathrm{Var}\left[\sum_{j=1}^{n-1} y_i y_{i+1}\right] &= \E\left[\left(\sum_{j=1}^{n-1} y_i y_{i+1}\right)^2\right]\\
&= \E\left[\sum_{h=1}^{n-1} \sum_{j=1}^{n-1} y_i y_{i+1} y_h y_{h+1}\right] &= \E\left[\sum_{j=1}^{n-1} y_i^2 y_{i+1}^2\right] = (n-1){\sigma}^4\,. \nonumber
\end{eqnarray}
For large $n$, the classical Central Limit Theorem shows that $\sum_{j=1}^{n-1} y_i y_{i+1}$ is asymptotically normal, distributed as $N(0,(n-1)\sigma^4)$. The denominator, when divided by $(n-2)$, is an estimator of the variance. Therefore
\begin{equation}
\hat{\rho}(1) = \frac{\sum_{j=1}^{n-1} y_i y_{i+1}}{\sum_{i=1}^n y_i^2} \sim N\left(0,\frac{1}{n}\right)\,,
\end{equation}
for $n >> 1$. For this reason the $95\%$ confidence interval of autocorrelation coefficient can be approximated by $\pm 2 \sqrt{1/n}$. The null hypothesis of non significance of a value is then based on the normality and stationarity of the increments. The first hypothesis can easily be weakened by simply imposing the finiteness of the second moment, but the second is much harder to do without, and our data are affected by non-stationarity. We will however rely on this confidence level to provide a better picture of the obtained results, but the reader should consider those confidence intervals with a somewhat suspicious mind.

\subsection{Hurst exponent}

The Hurst exponent is defined in the framework of fractional Brownian motion, a well-known stochastic process where
the second order moments of the increments scale as
\begin{equation}
\mathbf{E}\left[(x(t_2)-x(t_1))^2\right] \propto \left| t_2 - t_1\right|^{2H}\,,
\end{equation}
with $H \in [0, 1]$. The Brownian motion is the particular case when $H = 1/2$.
If $H < 1/2$ the behaviour of the process is anti-persistent, that is, deviations of one sign are generally followed by deviations with the opposite sign, an effect that in finance is usually called mean reversion. The limiting case $H = 0$, corresponds to white noise, where fluctuations at all frequencies are equally present or $1/f$-noise (pink noise), where the power spectral density is inversely proportional to the frequency.
If $H > 1/2$, then the behaviour of the process is persistent, i.e. consecutive increments
 tend to have the same sign and we observe a trend-following process. The limiting case $H = 1$, reflects $x(t) \propto t$ (locally), i.e. a smooth function.

As pointed out in the introduction, the correspondence between the values of $H$ and the behaviours described above is true only in the particular framework of the fractional Brownian motion. Otherwise, an estimation of $H$ is only an indicator of the scaling properties of the time series \cite{Embrechts2002}. As a limit example, one can consider the case of the stable L\'evy processes. A stability index $\alpha$ leads to a Hurst exponent equal to $\max(1,1/\alpha)$, but there is no presence of persistent behaviour.

Therefore, we will not rely on the mere estimation of the Hurst exponent to assess whether our final time series is efficient or not, but we will rather use it as an indicator of a possible special behaviour.

\section{The procedure}
We now describe our method to find the suitable weights $w_i$ leading to the desired ``non-efficient'' basket. Since we have decided to focus our attention on negative autocorrelation, we will start looking for the basket that minimize this statistic in a fixed period of time. More precisely, we consider a number of consecutive days $D$ and we look for the set of weights producing $B_D$, the basket with the minimum value of autocorrelation possibly obtainable given the recorded data. Please beware that this is an \emph{a posteriori} evaluation: we first observe the data, and then choose the weights. At this stage, the efficiency of the market is not under attack. The efficiency is beaten only if we are able to provide the weights of a negatively autocorrelated basket without using the information of the data we are testing. And to do so, immediately after the $D$ consecutive days, that we call \emph{minimization period}, we pick $S$ consecutive days in which we test $B_D$. If the value of the anti-autocorrelation remains significantly low during those $S$ days, we now have evidence for non-efficiency.

We can choose among a plethora of methods to optimize the property we are interested in \cite{Ciarlet1989,Schneider2006}. For the sake of simplicity, we will use the autocorrelation matrix. This approach presents some difficulties: for example we are working with scales in which the Epps effect is strong \cite{Epps1979,Toth2009}, and, aside of this, we will have to estimate $N\times N$ parameters, each one effected by a measurement error. Aside from these purely technical aspects, we could also define other measures of mean-reversion than the autocorrelation. For example the sign correlation, or the Hurst exponent itself (just remembering the interpretation problems quickly explained in Sec.~\ref{sec:introduction} and \ref{sec:autocorrelation}), would be reasonable choices, and in those cases the matrix approach would not work, or at least not in its usual formulation.

The classical correlation matrix approach to collective behaviour involve equal
time correlations. Instead, the construction of a delay correlation matrix (or time lagged correlation matrix) involves calculating correlations between different assets fixing a time delay \cite{Biely2008,Mayya2007}. 
Let us define a matrix $\mathbf{M}$ of order $N\times T$ where each row is filled by the records of a discretized time series; $N$ is the number of time series of length $T$. Suppose $i$ and $j$ are two distinct assets among the given time series. The delay correlation matrices between $i$ and $j$ at time lag $k$ is given by
\begin{equation}
\mathbf{\hat{C}}_{i,j} = \frac{1}{T} \sum_{t=1}^{T-k} \mathbf{M}_{i,t} \mathbf{M}_{j,t+k}\,
\end{equation}
where with $\mathbf{A}_{h,k}$ stands for the element in the $h^{\mathrm{th}}$ row and $k^{\mathrm{th}}$ column of the $\mathbf{A}$ matrix.
The matrix $\mathbf{\hat{C}}$ thus constructed is asymmetric, but the associated quadratic minimization problem $\mathbf{e}^\mathsf{T} \mathbf{\hat{C}} \mathbf{e}$ is equivalent to that in the symmetrized case with corresponding matrix
\begin{equation}
\mathbf{C}_{i,j} = \mathbf{C}_{i,j} = \frac{\mathbf{\hat{C}}_{i,j} + \mathbf{\hat{C}}_{j,i}}{2}\,, 
\end{equation}
and it can be simply justified observing that $(\mathbf{e}^\mathsf{T} \mathbf{\hat{C}} \mathbf{e})^\mathsf{T} = \mathbf{e}^\mathsf{T} \mathbf{\hat{C}}^\mathsf{T} \mathbf{e}$. The minimum eigenvalue of this matrix represents the mode for which the auto-correlation at time lag $K$ is minimized. This means that fixing $k=1$, filling the matrix $\mathbf{M}$ with the increments of the analyzed stocks during the $D$ minimization days (increments normalized by the sample volatility in this period) and finding the eigenspace corresponding to the minimum eigenvalue, we have found the basket with the minimal possible autocorrelation, among all baskets with unit euclidean norm; $w_i$ must be equated to the $i^{\mathrm{th}}$ component of the eigenvector. So, calling these components $e_1,\ldots,e_N$ we obtain
\begin{equation}
B_{Dt} = \sum_{i=1}^N e_i x_{it}
\end{equation}
where $x_{it}$ is the element of the $i^{\mathrm{th}}$ discretized time series and at discrete time $t$. Please note that $x_{it}$ are the \emph{raw} increments and they are not normalized by the variance as into the matrix $\mathbf{M}$.

The minimum eigenvalue by itself does not give us the value of the autocorrelation; we must compute it directly from $B_D$ using the estimator in Eq.~(\ref{eq:autocorrest}) and we will repeat this operation in both the minimization period and the subsequent test period.

So, schematically, we proceed as follow:
\begin{enumerate}
\item We consider $D$ consecutive days and the immediately following $S$ days.
\item The time series are discretized (Sec.~\ref{sec:data}), normalized by the variance during the $D$ days, and used to build the matrix $\mathbf{C}$ with $k=1$.
\item From the components of the eigenvector corresponding to the smallest eigenvalue we can build the basket $B_D$.
\item We evaluate the autocorrelation of $B_D$ during the whole minimization period and, independently, during the test period.
\item Leaving the number of days $D$ fixed we shift the minimization period ahead by one day, and we go back to point 2.
\end{enumerate}

The considered time scales (the discretization time step) are $5,10,\ldots,55,60$ seconds.

The number of test days $S$ is arbitrary. Since we are fixing $B_D$ outside these data we must only take care of having enough statistics but we cannot increase much the minimization period, for fear of structural changes in the inter-dependences between the stocks that could cancel the desired effect. Therefore we will test $B_D$ for $S=1$ and $S=5$.

The choice of $D$ is more tedious. By choosing a small number of days, we decrease the size of the data set and we are more likely to catch some unstable interdependences, where by unstable we mean dependences standing only for a short period of time. On the other hand, considering a large number of days $D$ we take the risk of averaging out the time changes, finding at the end a non satisfactory value of the autocorrelation. As trade off we fix $D=10$.

Tab.~\ref{tab:popolazione} contains the length of the time series when these numbers are applied.

In addition, we compute the Hurst exponent of $B_D$ using the periodogram method. As stated above, the Hurst exponent alone does not give any information regarding the persistence of the autocorrelation and, moreover, the estimators are well known to be inaccurate (biased) in most of the situations. For those reason we do not use more complex estimators, nor do we want to speculate on the values we find, but rather present them only as indication of ``possible'' persistence effects. The estimation of the Hurst exponent is carried out each time considering a time scale of $5s$. This is because the Hurst exponent measures a global property of the process; stretching the time scale would only decrease the statistics and cutting the effects at short time scales.
\begin{figure}
\centering
\includegraphics[scale=0.5]{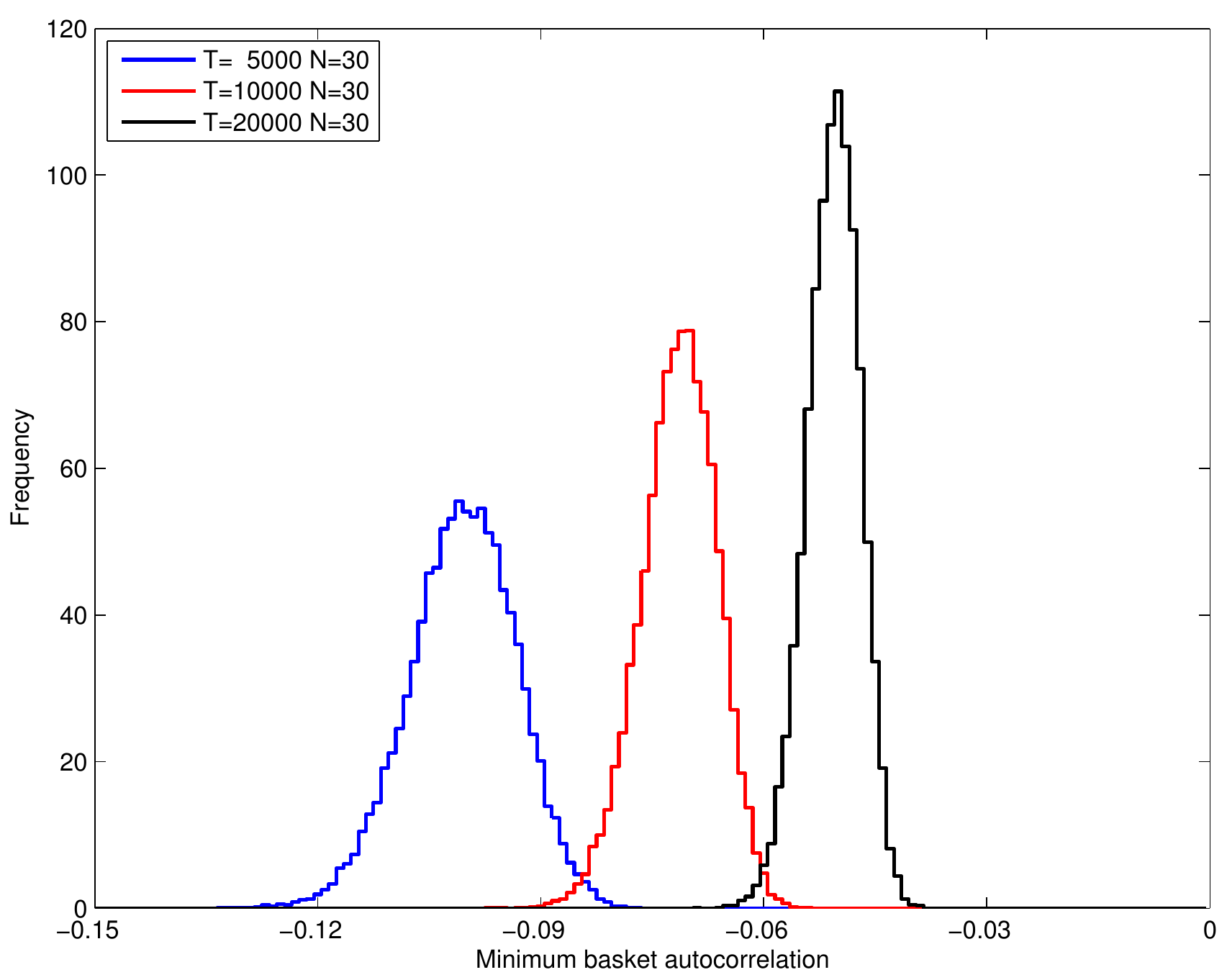}
\caption{Empirical distribution of the optimal anti-autocorrelation for a basket
build up of 30 synthetic stock timeseries with IID normally distributed
increments.}
\label{fig:white}
\end{figure}

\begin{table}
\begin{center}
\begin{tabular}{cccc}
     & Minimization $D=10$ & Test $S=1$ & Test $S=5$\\
\hline
5 s  & 41400		& 4140       & 20700 \\
10 s & 20700		& 2070		 & 10450 \\
15 s & 13800		& 1380		 & 6900 \\
20 s & 10350		& 1035       & 5175 \\
25 s & 8280			& 828		 & 4140 \\
30 s & 6900			& 690		 & 3450 \\
35 s & 5910*		& 591       &  2455 \\
40 s & 5170*		& 517		 & 2065 \\
45 s & 4600			& 460		 & 2300 \\
50 s & 4140			& 414       &  2070 \\
55 s & 3760*		& 376		 & 1880 \\
60 s & 3450			& 345		 & 1525 \\
\end{tabular}
\end{center}
\caption{Length of the time series as function of the considered time window size. For the starred values the considered time could have not been extended precisely until 3.45 pm; in these cases the considered period cannot be perfectly divided into the desired time steps, so some seconds preceding 3.45 pm have been discarded. }
\label{tab:popolazione}
\end{table}

The autocorrelations values for the out-of-sample periods can be tested using the confidence interval discussed in section Sec.~\ref{sec:autocorrelation}. On the other hand, for the values obtained in the minimization periods we cannot follow this simple rule. They are the smallest values possibly obtainable by the given data so the null hypothesis cannot be the absence of autocorrelation. It must rather be the likelihood to obtain such values as minimization of independent time series. A heuristic approach is to run the minimization procedure described above in this section using a set of $N=30$ synthetic stocks with IID, normally distributed increments and zero correlations. Doing so, we are able to plot the distribution of the minimal values obtainable by simulation. In Fig.~\ref{fig:white} this distribution is reported for the values of $T=5000,10000,20000$ and for a statistic made by a population of $20000$ synthetic baskets. Even for the shorter sample (small $T$), where spurious effects are more likely to occur, the value is hardly larger than $0.12$, and the typical values drop sensitively for the larger samples.

\begin{figure*}
\centering
\includegraphics[height=5.5cm, width=.49\columnwidth]{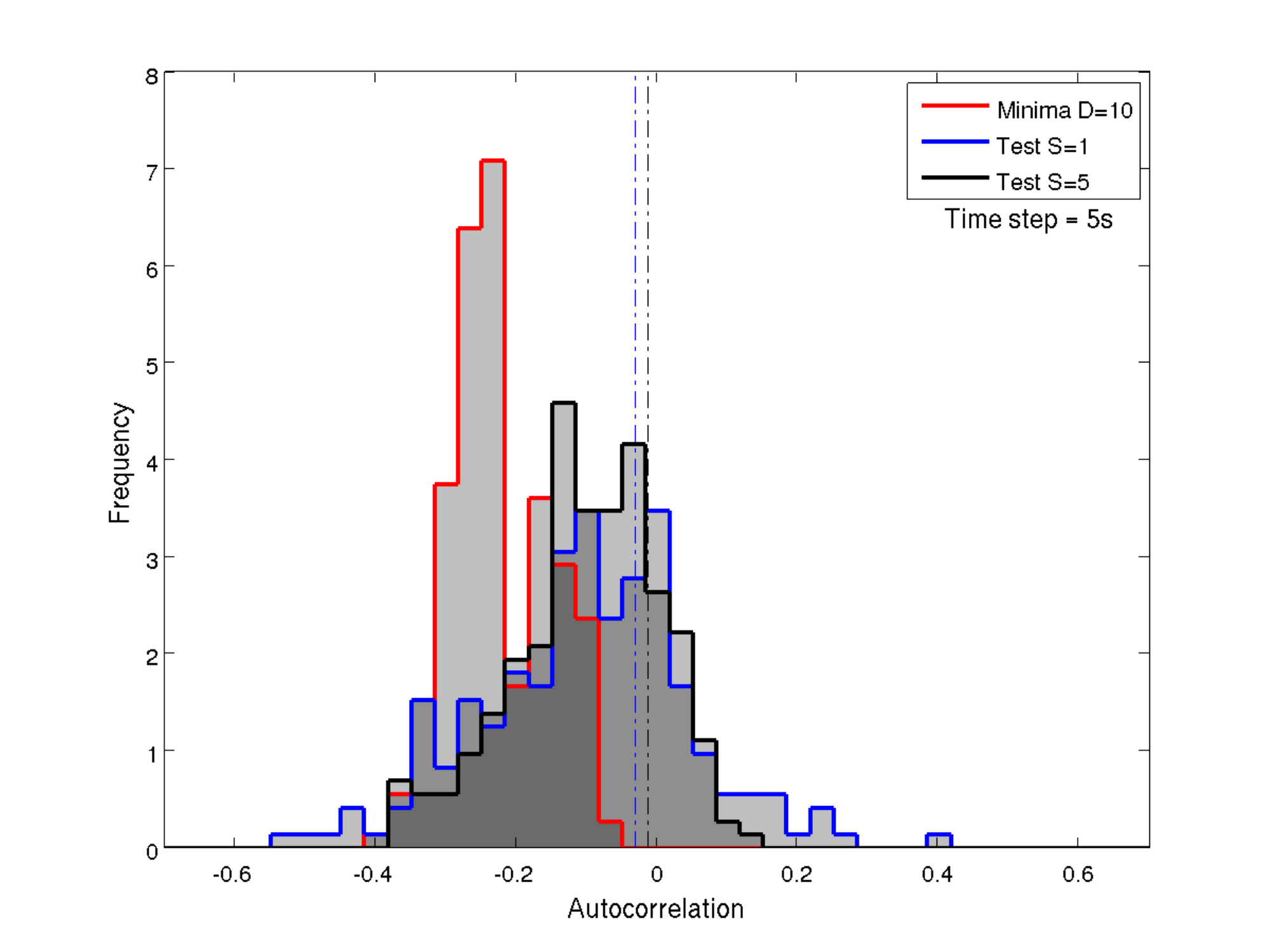}
\includegraphics[height=5.5cm, width=.49\columnwidth]{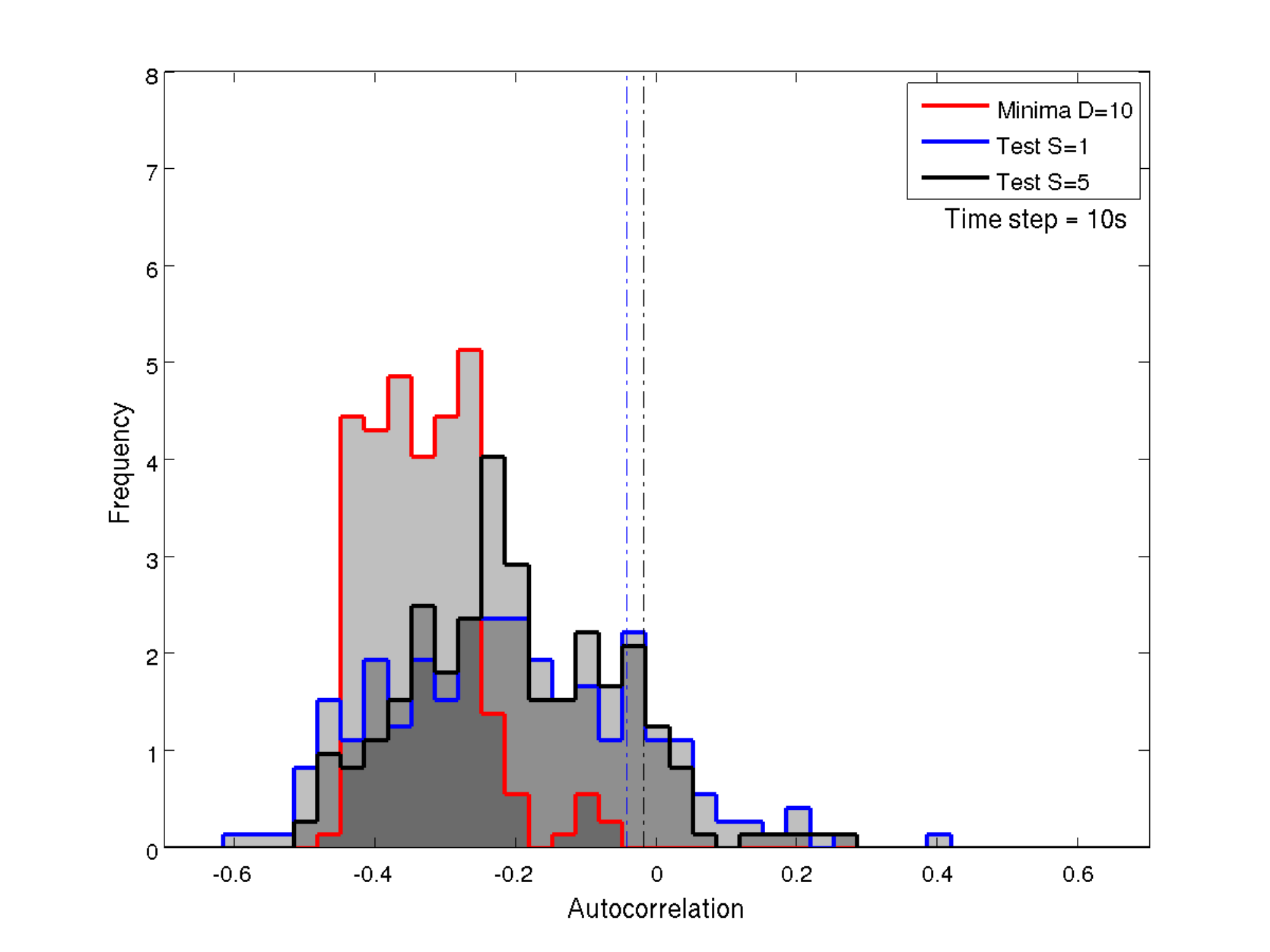}\\
\includegraphics[height=5.5cm, width=.49\columnwidth]{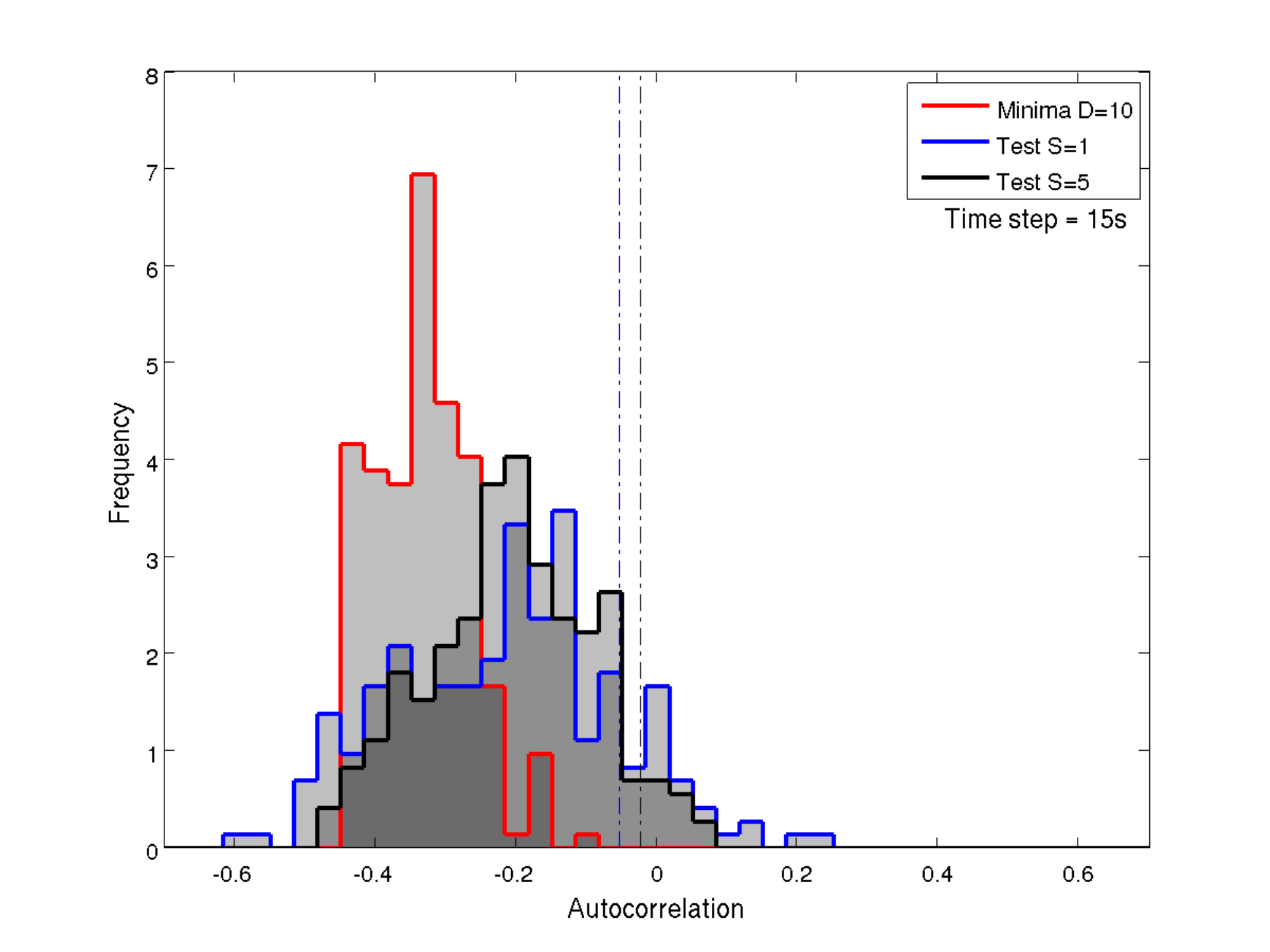}
\includegraphics[height=5.5cm, width=.49\columnwidth]{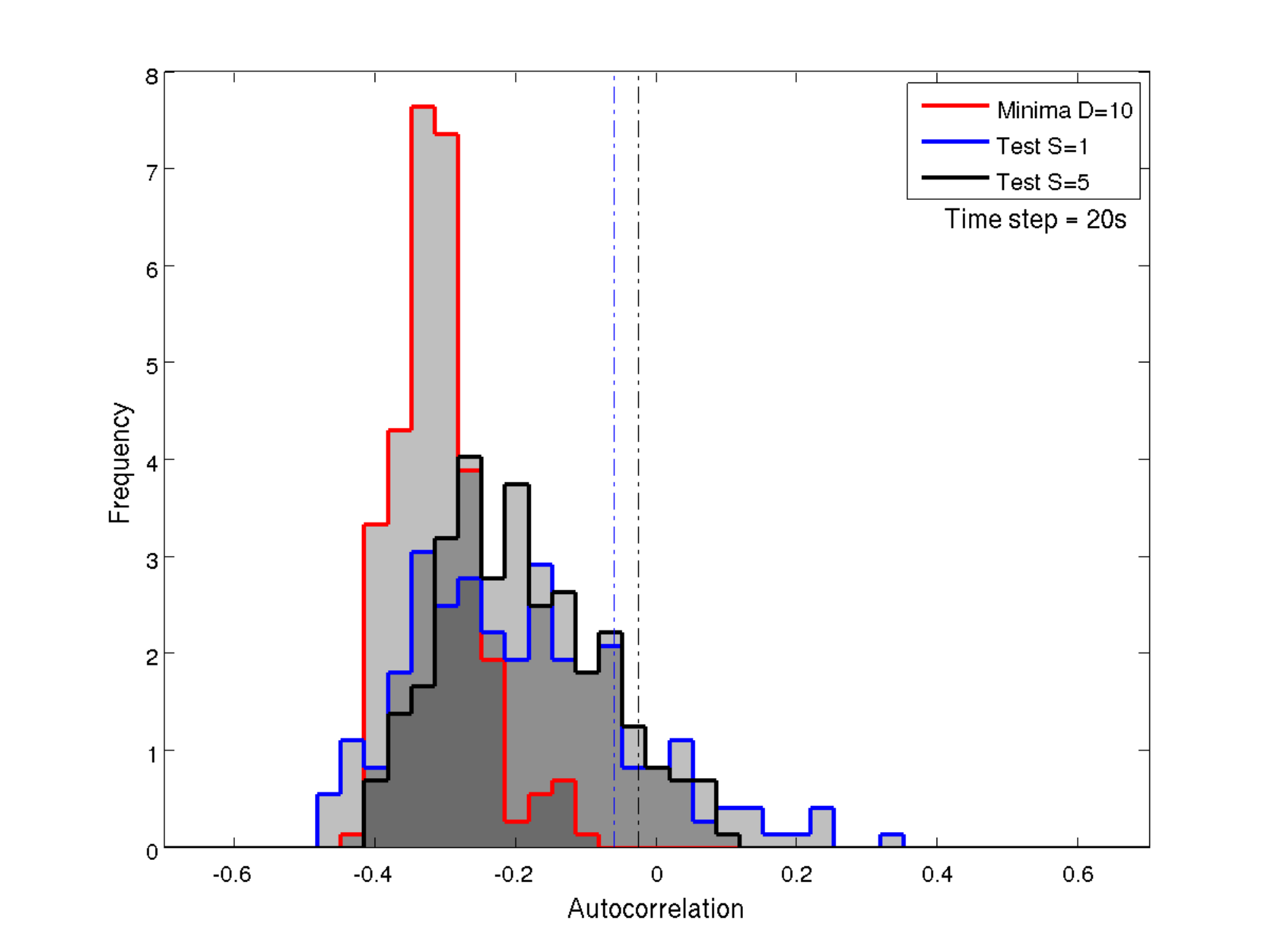}\\
\includegraphics[height=5.5cm, width=.49\columnwidth]{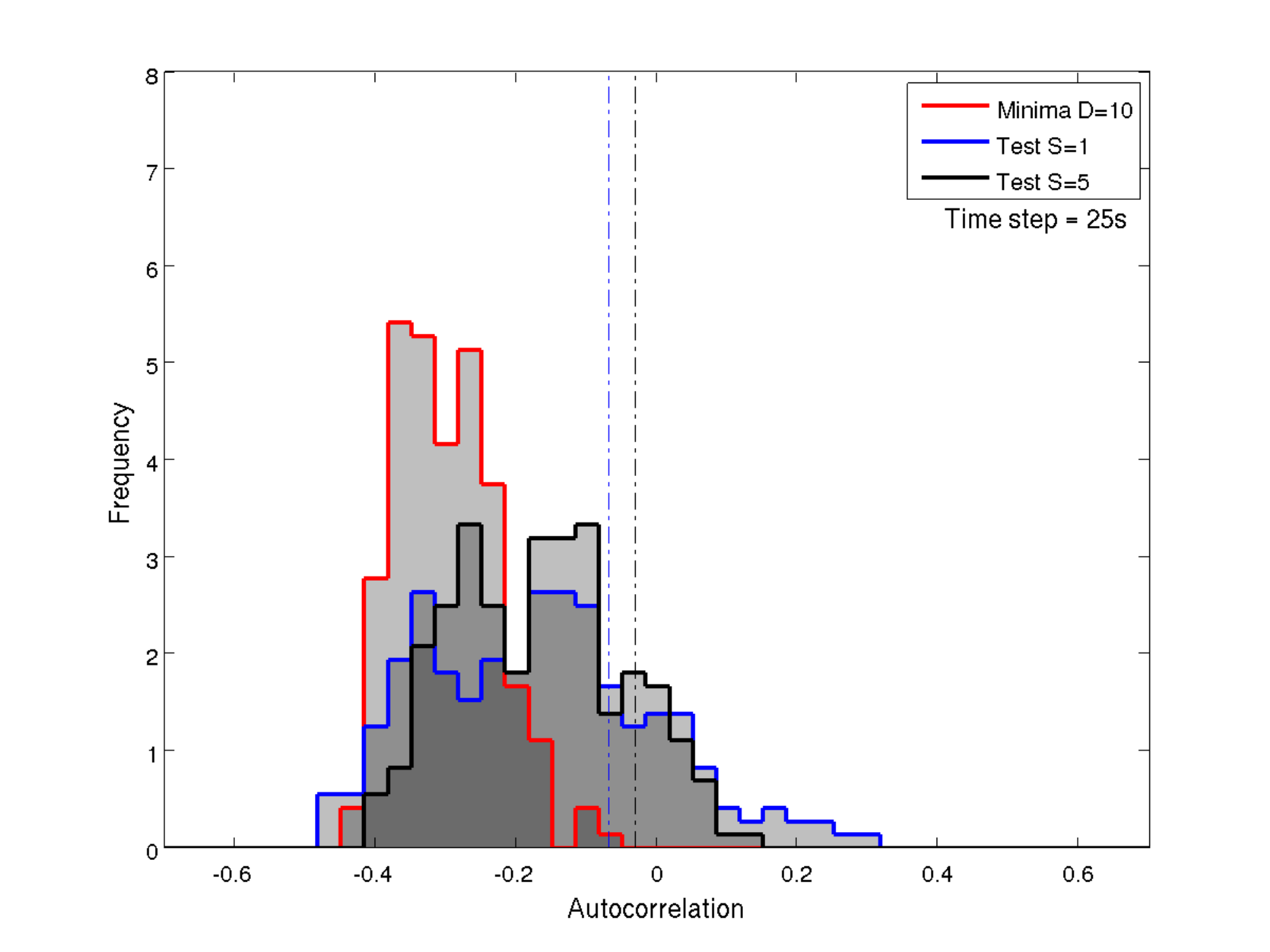}
\includegraphics[height=5.5cm, width=.49\columnwidth]{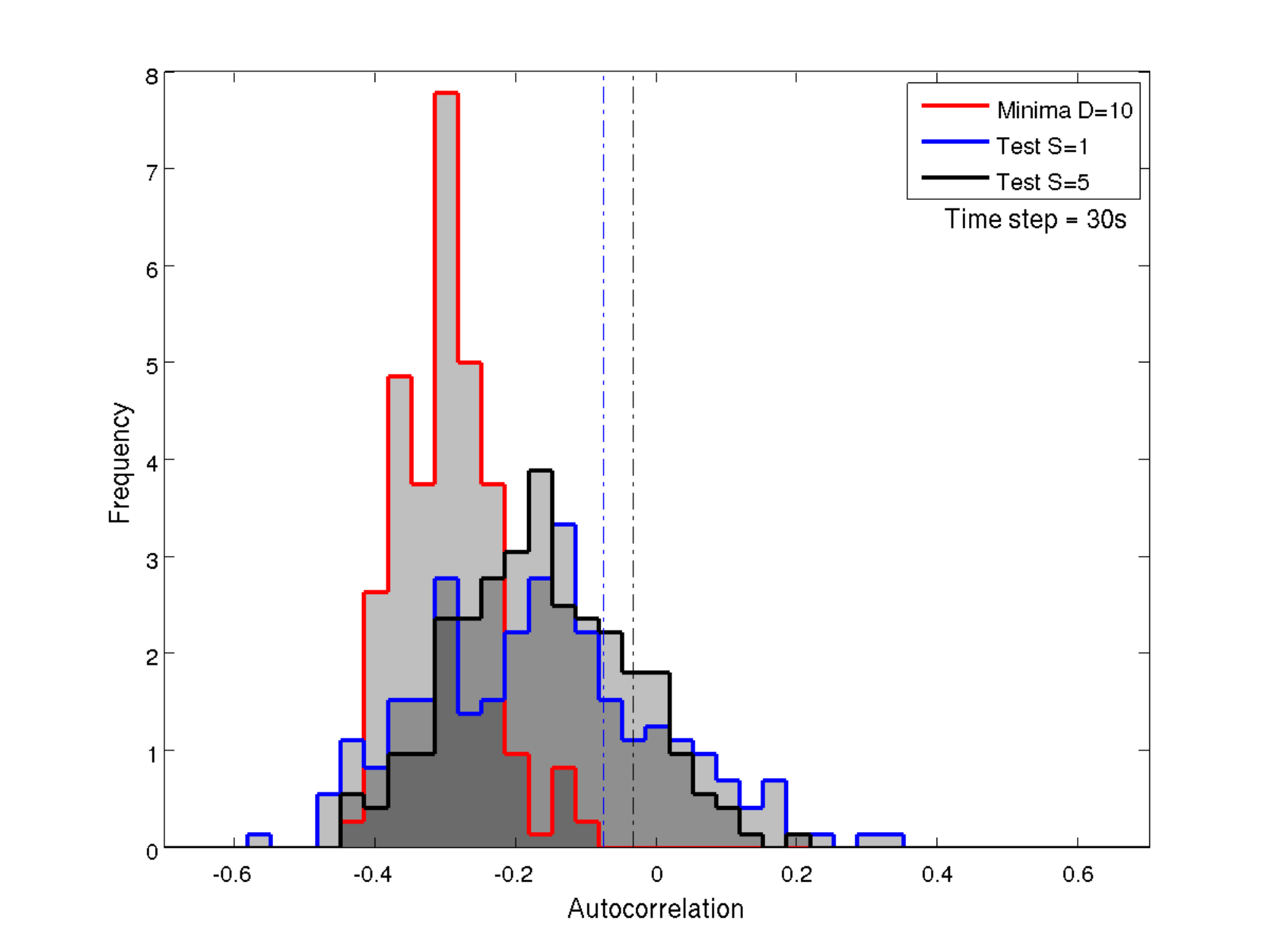}\\
\caption{Histograms of the resulting autocorrelation values. In red we have the distributions for the minimal values obtained during $D=10$ consecutive days, in blue and black we have the distributions for the test performed considering respectively $S=1$ and $S=5$ subsequent days. The dashed lines are the critical values for the significance test using the 5\% of confidence.}
\label{fig:res1}
\end{figure*}

\begin{figure*}
\centering
\includegraphics[height=5.5cm, width=.49\columnwidth]{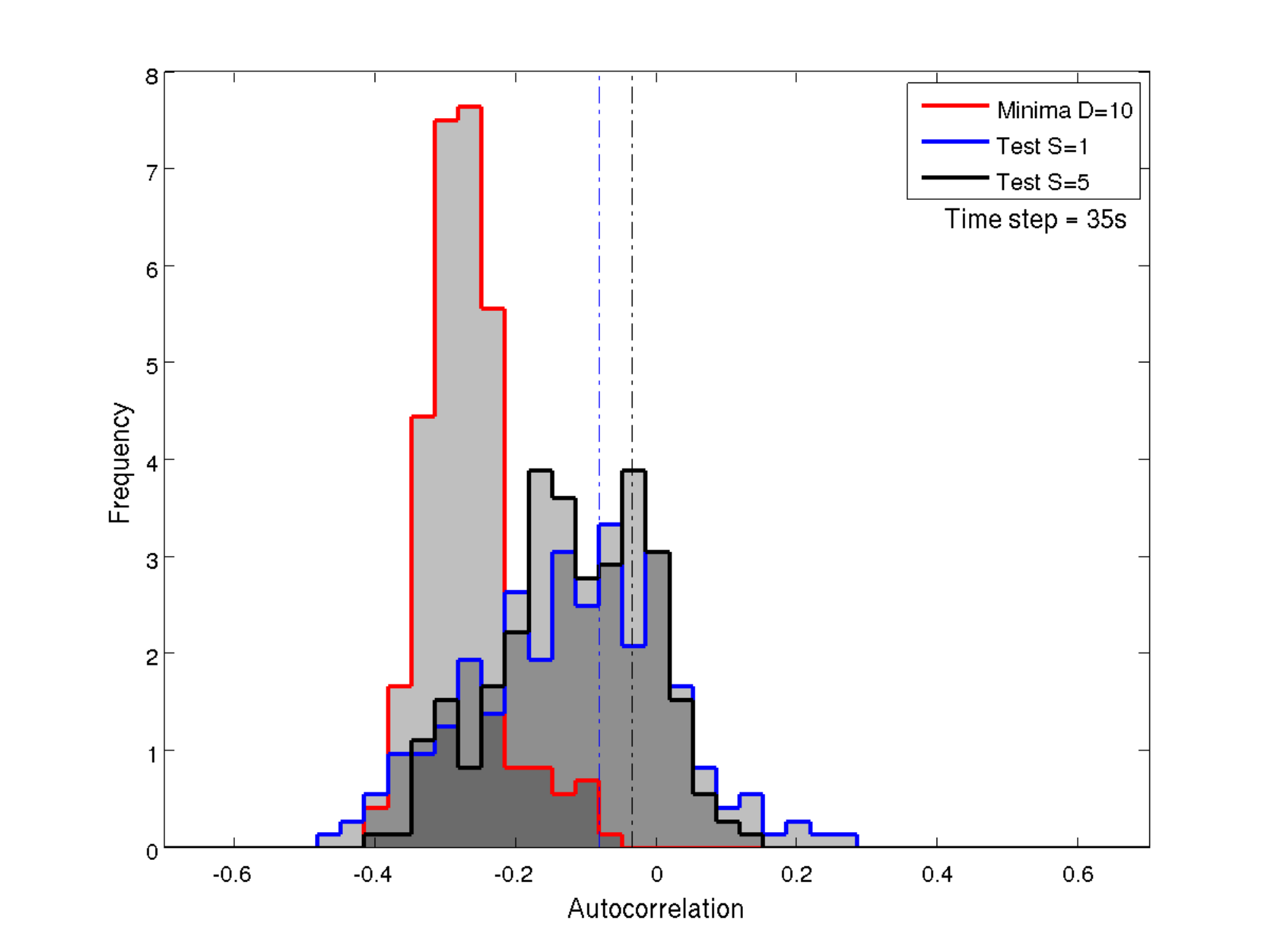}
\includegraphics[height=5.5cm, width=.49\columnwidth]{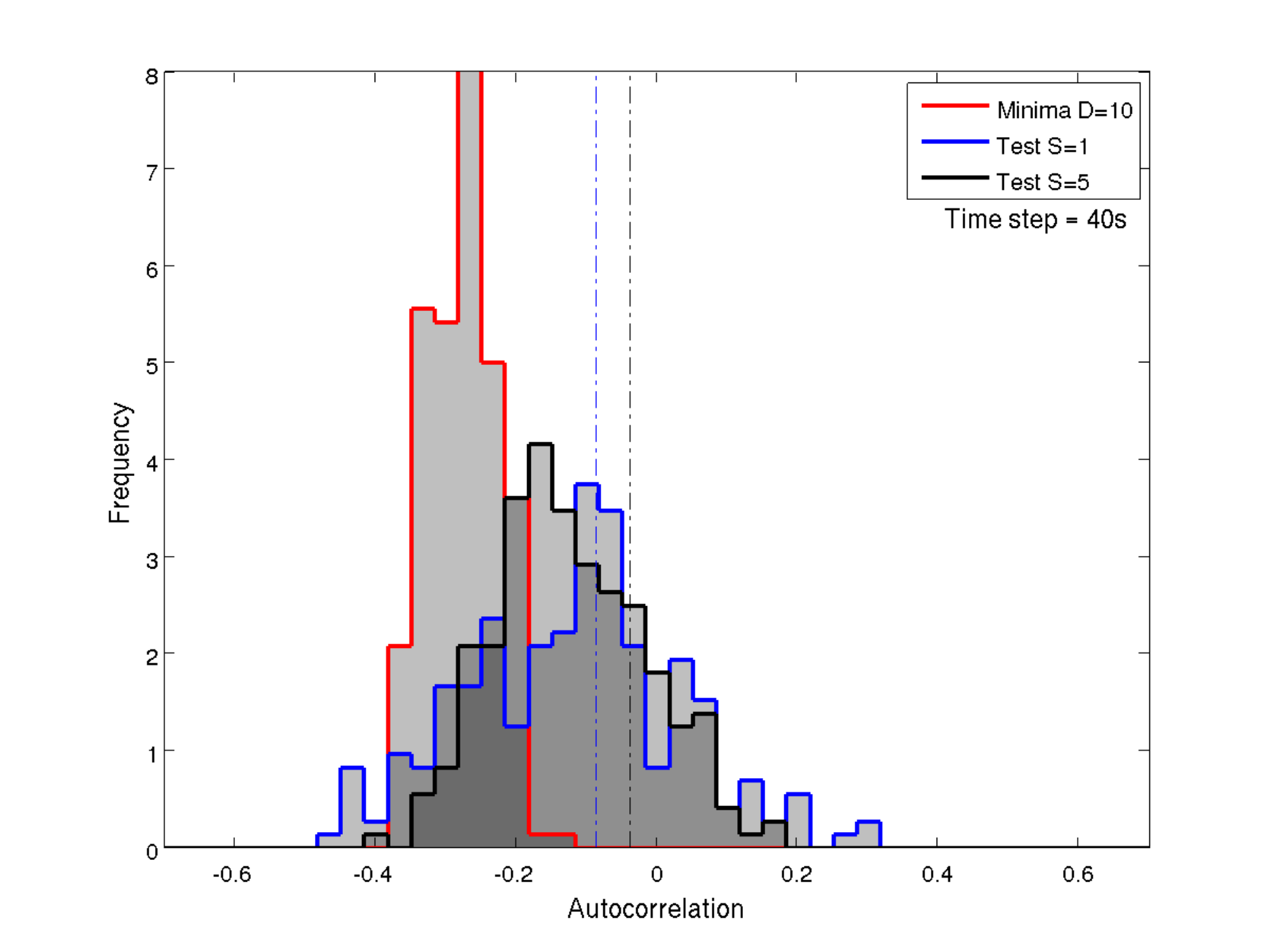}\\
\includegraphics[height=5.5cm, width=.49\columnwidth]{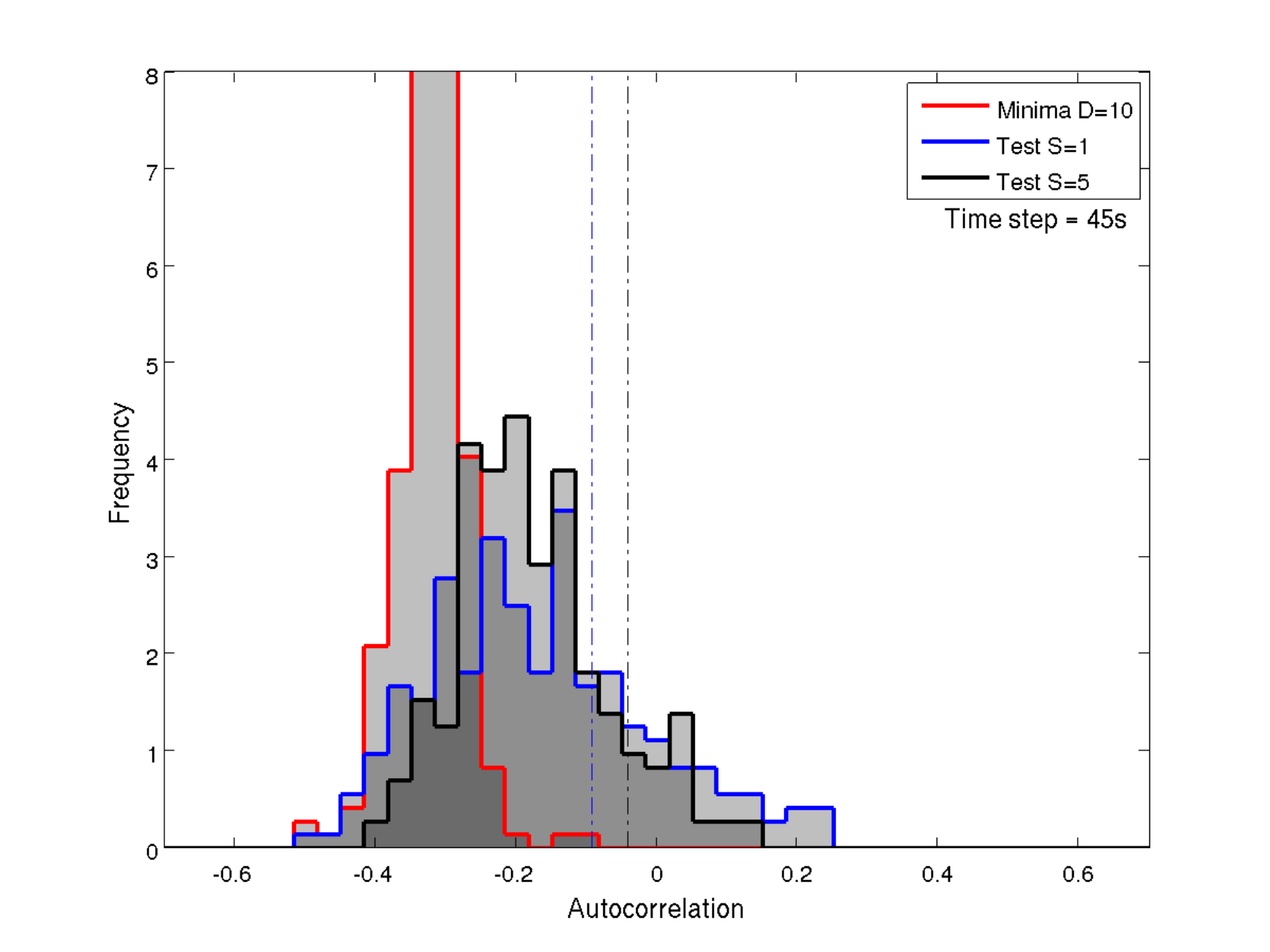}
\includegraphics[height=5.5cm, width=.49\columnwidth]{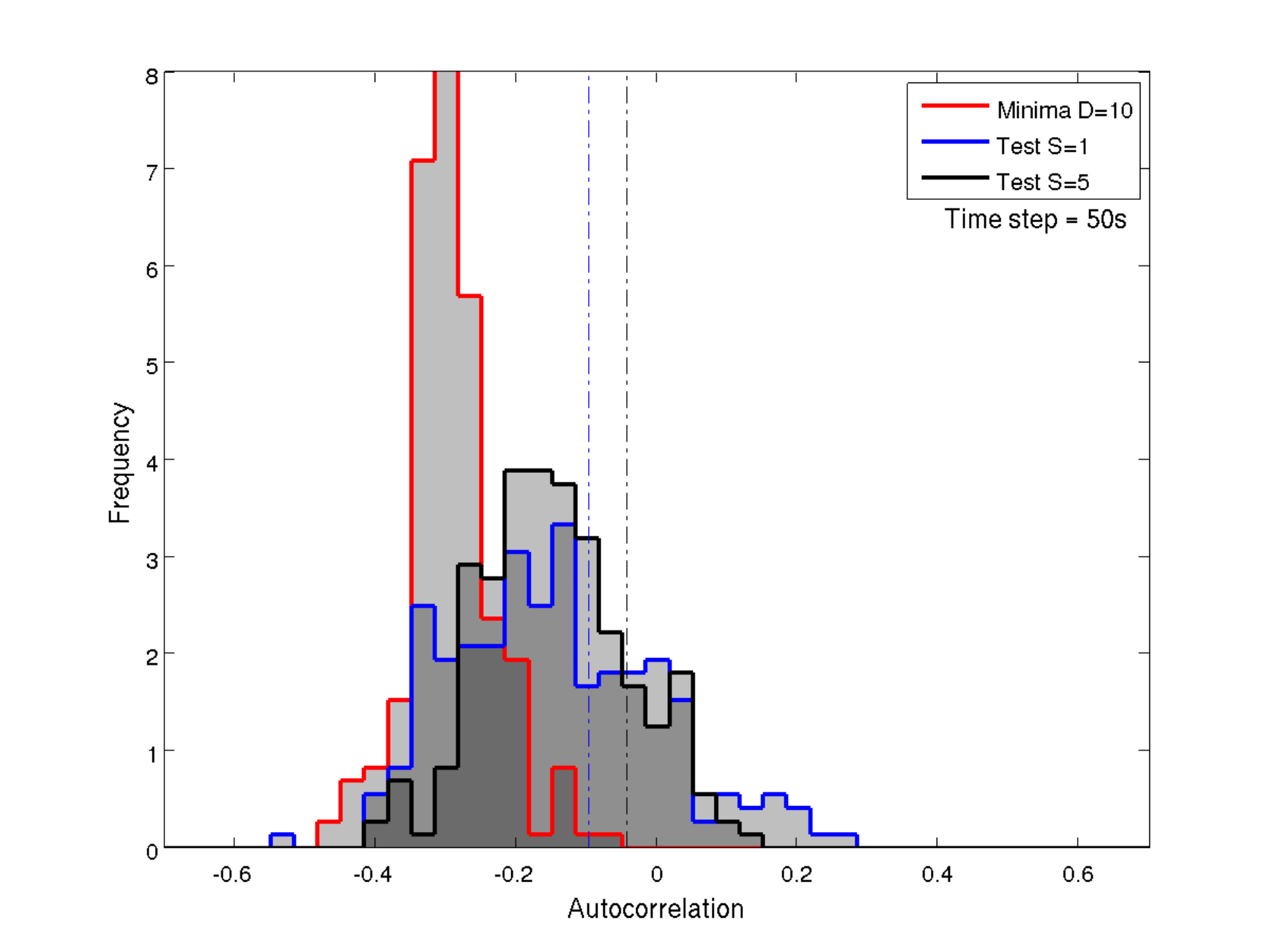}\\
\includegraphics[height=5.5cm, width=.49\columnwidth]{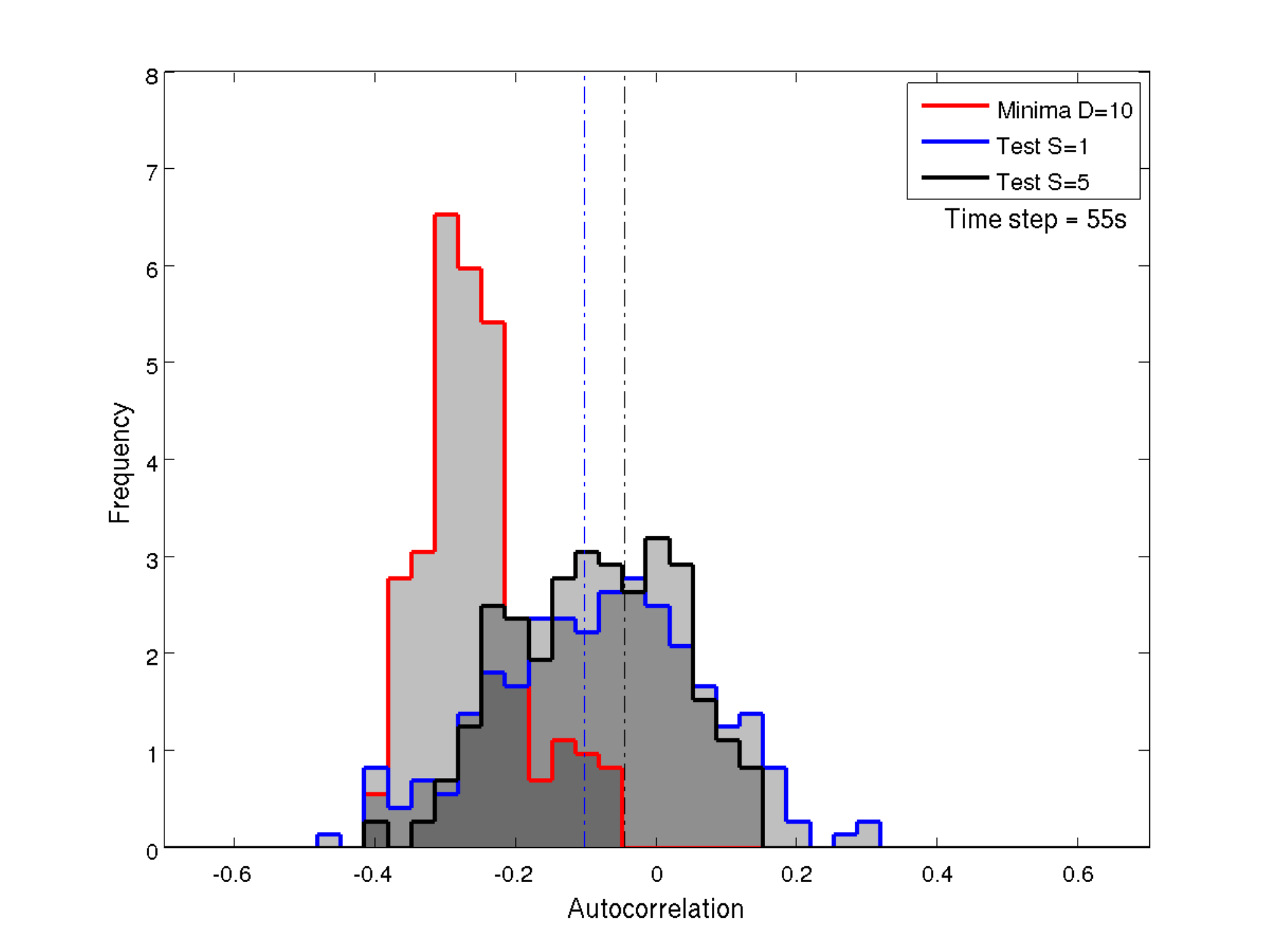}
\includegraphics[height=5.5cm, width=.49\columnwidth]{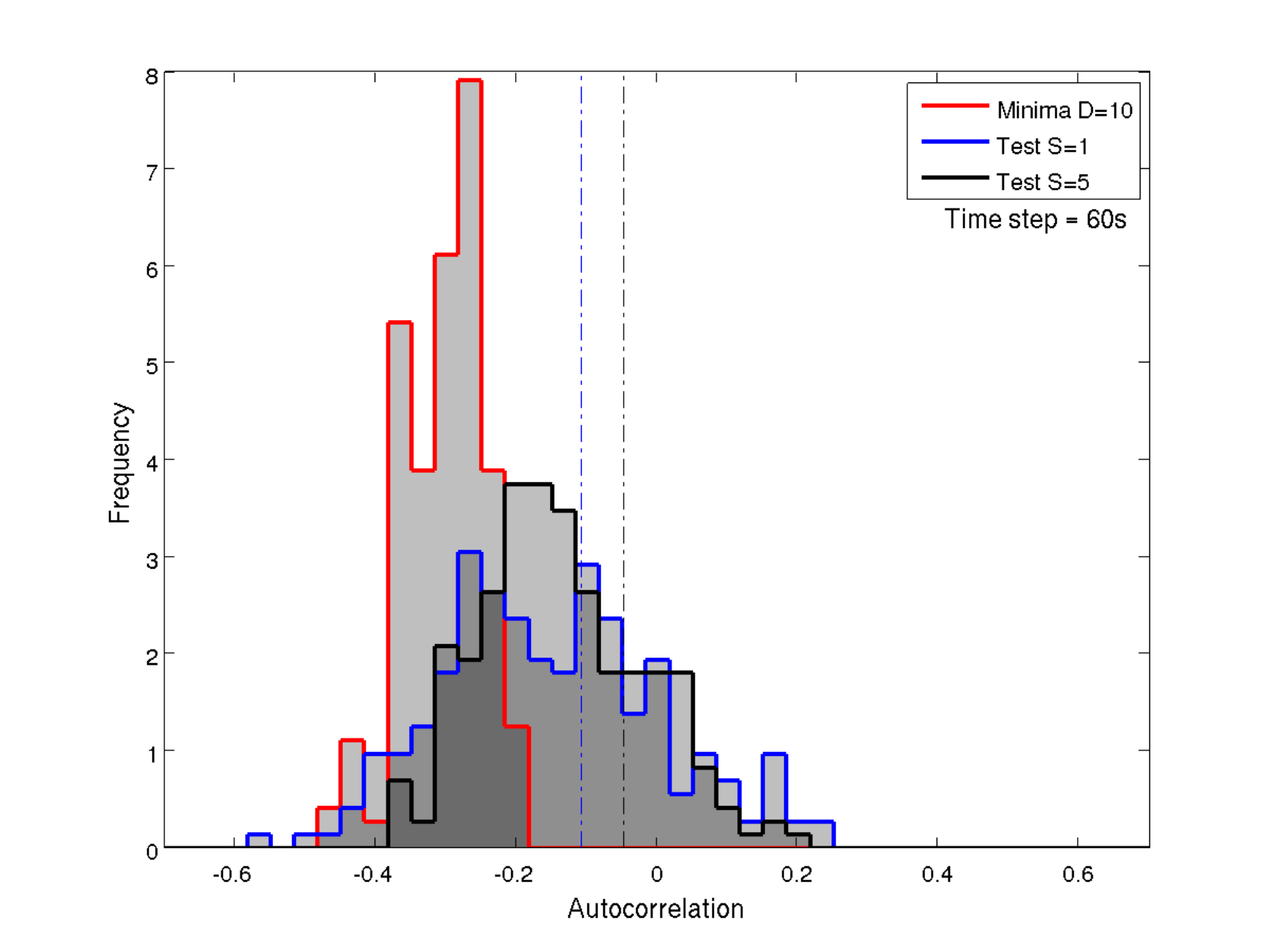}\\
\caption{Histograms of the resulting autocorrelation values. In red we have the distributions for the minimal values obtained during $D=10$ consecutive days, in blue and black we have respectively the distributions for the test performed considering $S=1$ or $S=5$ subsequent days. The dashed lines are the critical values for the significance test using the 5\% of confidence.}
\label{fig:res2}
\end{figure*}

\begin{figure}
\centering
\includegraphics[height=6.5cm, width=.95\columnwidth]{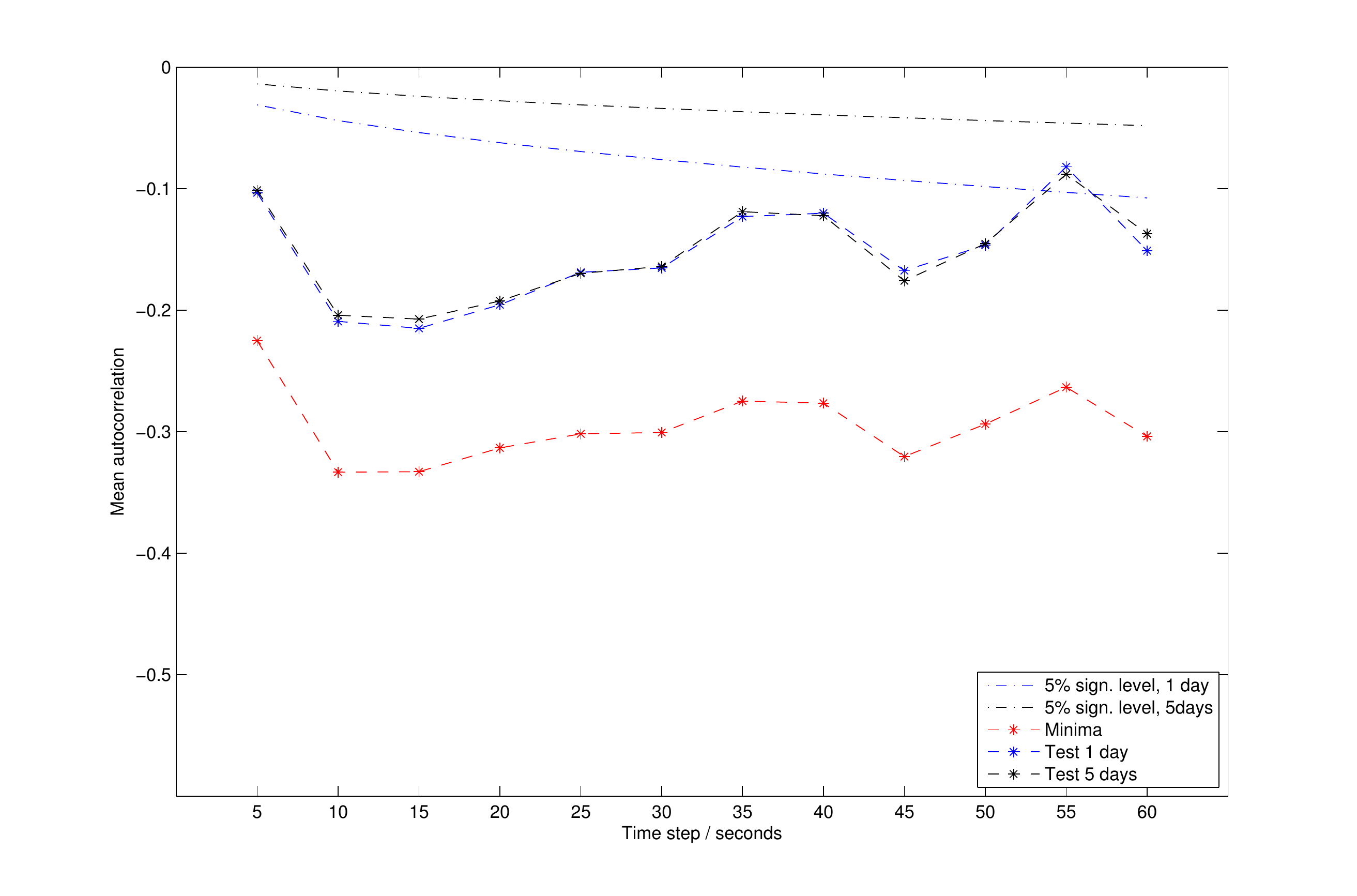}
\caption{Mean values of the samples as function of the considered time step. The red line stands for the minimized value of the initial sample, the blue line stands for the out-of-sample test using one day of data and, finally, the black line stands for the out-of-sample test using 5 days of data. At the top, the thin dashed lines (blue and black) reports the values of $5\%$ confidence level null hypothesis of non significant anticorrelation.}
\label{fig:meanvalues}
\end{figure}

\begin{figure}
\centering
\includegraphics[height=6.5cm, width=.95\columnwidth]{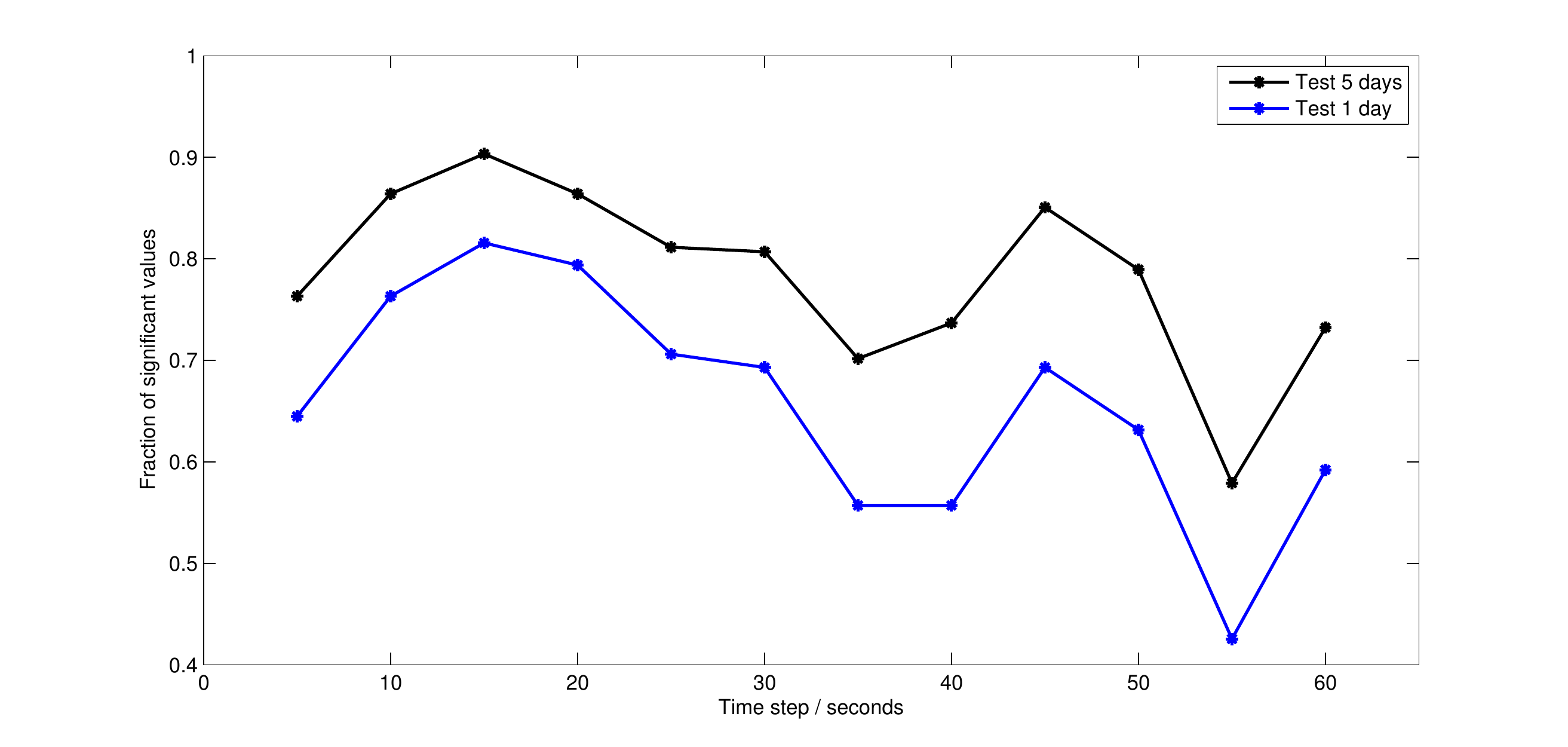}
\caption{Fraction of the out-of-sample test values that are rejected by the significance test, using the $5 \%$ confidence level. Blue stands for the one-day test and black for the five-day test. }
\label{eq:significative}
\end{figure}

\begin{figure}
\centering
\includegraphics[height=6.5cm, width=.95\columnwidth]{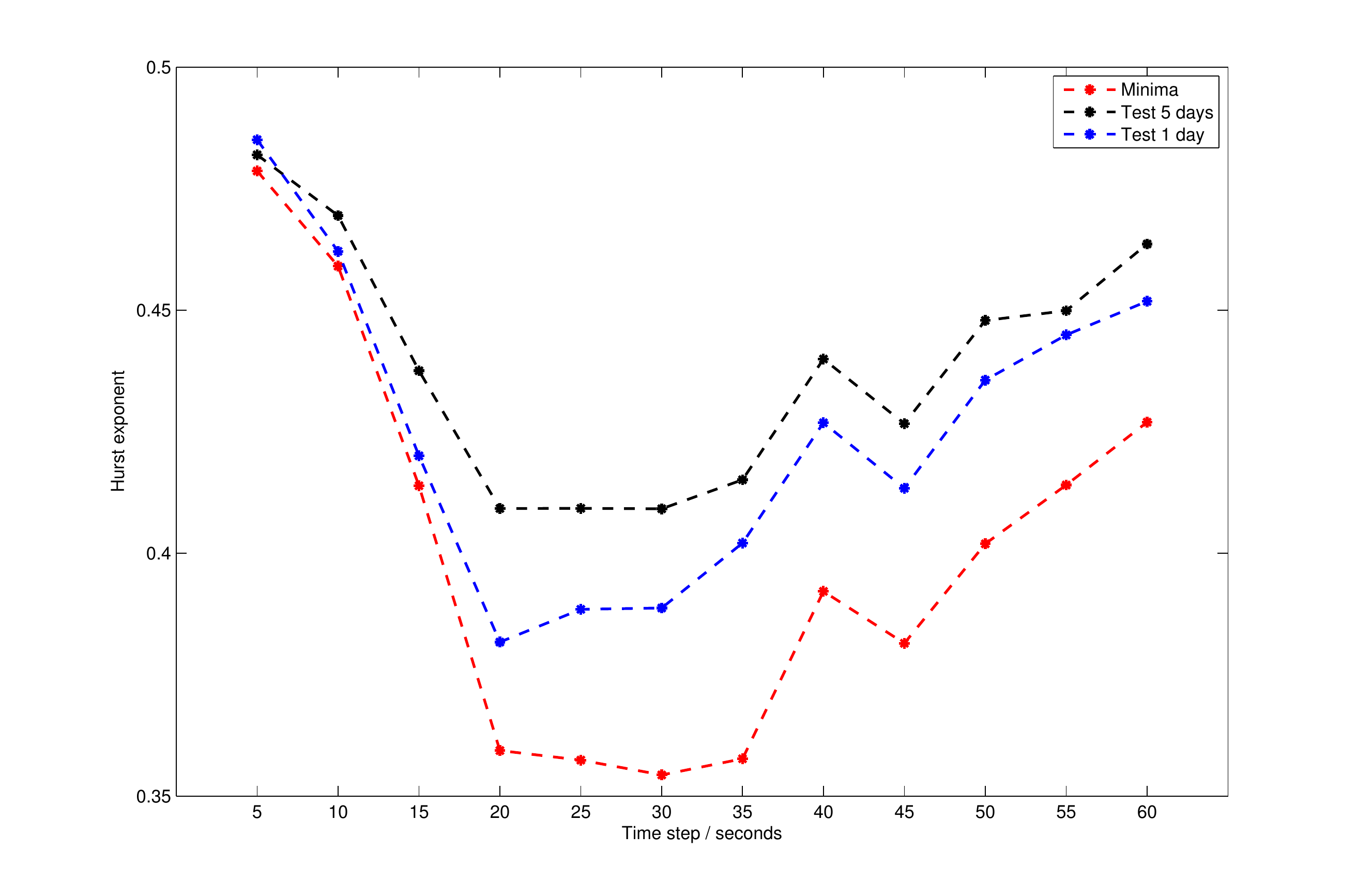}
\caption{Mean value of the Hurst exponent on the three samples. The minimized portfolio (red) presents a strongest mean Hurst exponent in the range 20~35s. The out-of-sample test using five days (black) still retains a significant value and presents the stronger effect around the scale 20~30s. }
\label{fig:hurst_results}
\end{figure}

\begin{figure}
\centering
\includegraphics[scale=0.5]{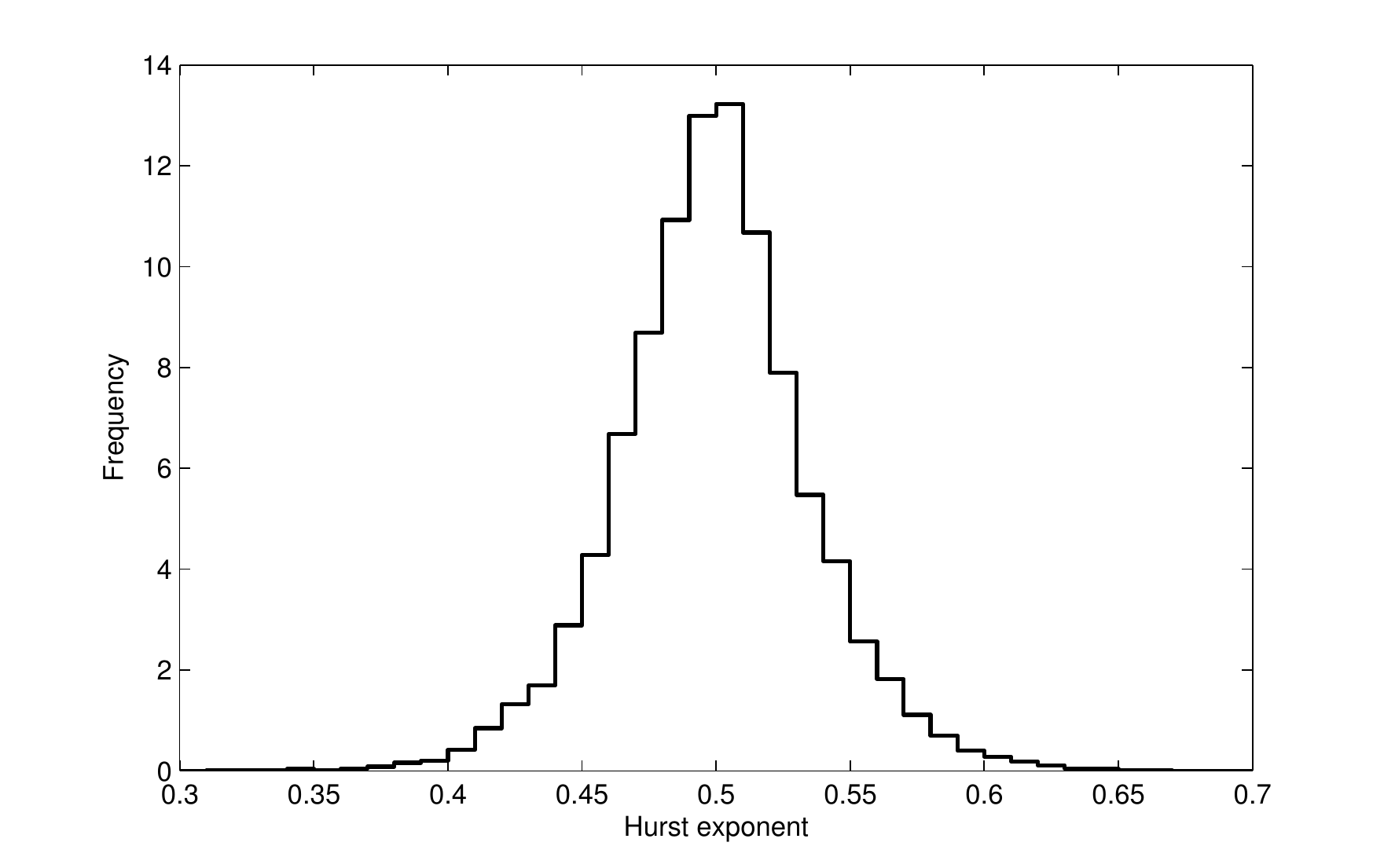}
\caption{Distribution of Hurst exponent estimated using the periodogram method by randomly extracting ten consecutive day of the DJIA data and building a basket. The population is made by $20000$ extractions and the mean value is $0.494$.}
\label{fig:hurst_rand}
\end{figure}

\begin{figure}
\centering
\includegraphics[scale=0.5]{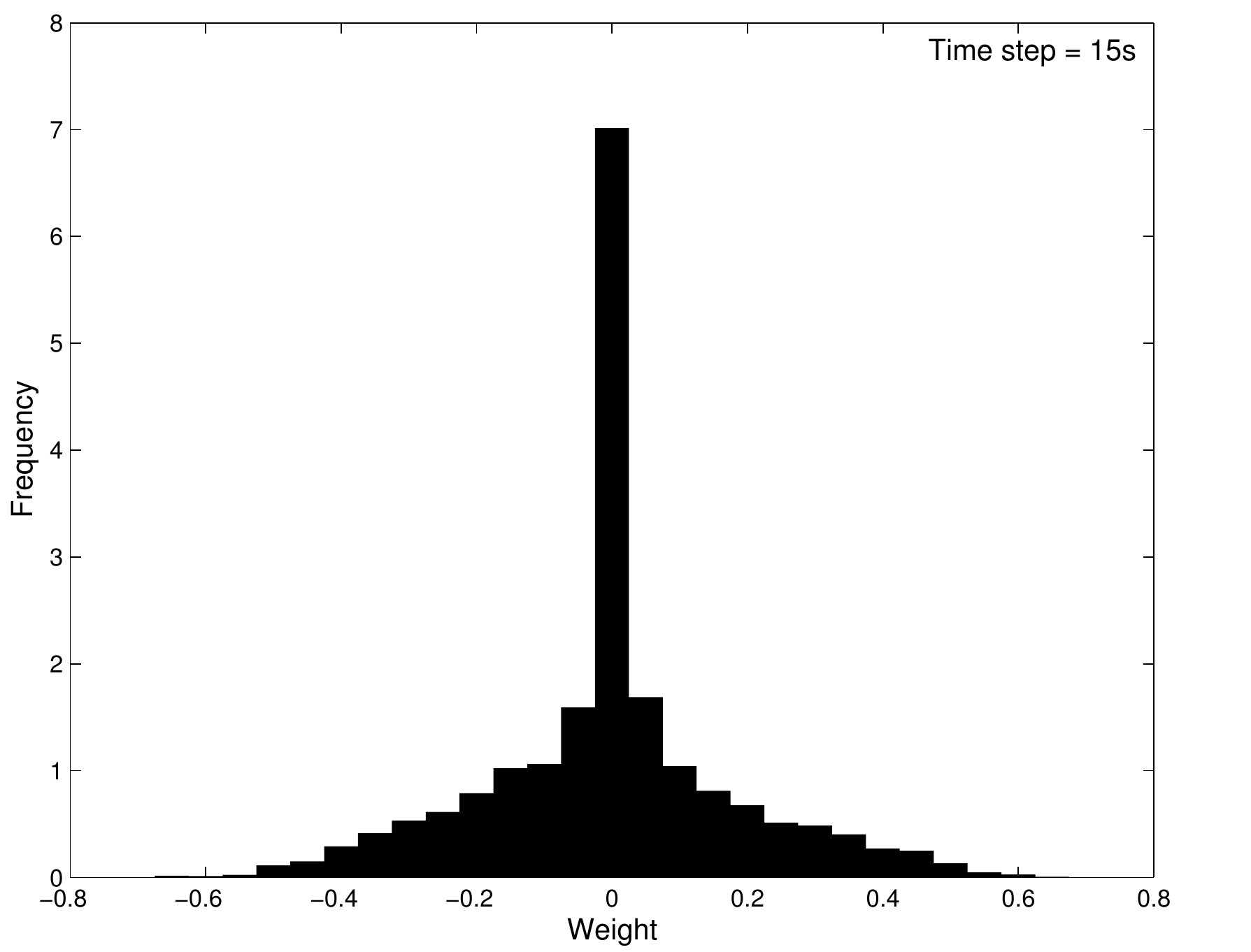}
\caption{Distribution of weights considering the unitarian Euclidean norm. The mass around zero is important but yet the distribution suggests the dynamics is actively participated by a large portion of the considered stocks. The depicted case corresponds to a time scale of $15s$; the remaining ones keep qualitatively the same look.}
\label{fig:weights}
\end{figure}

\begin{figure}
\centering
\includegraphics[scale=0.5]{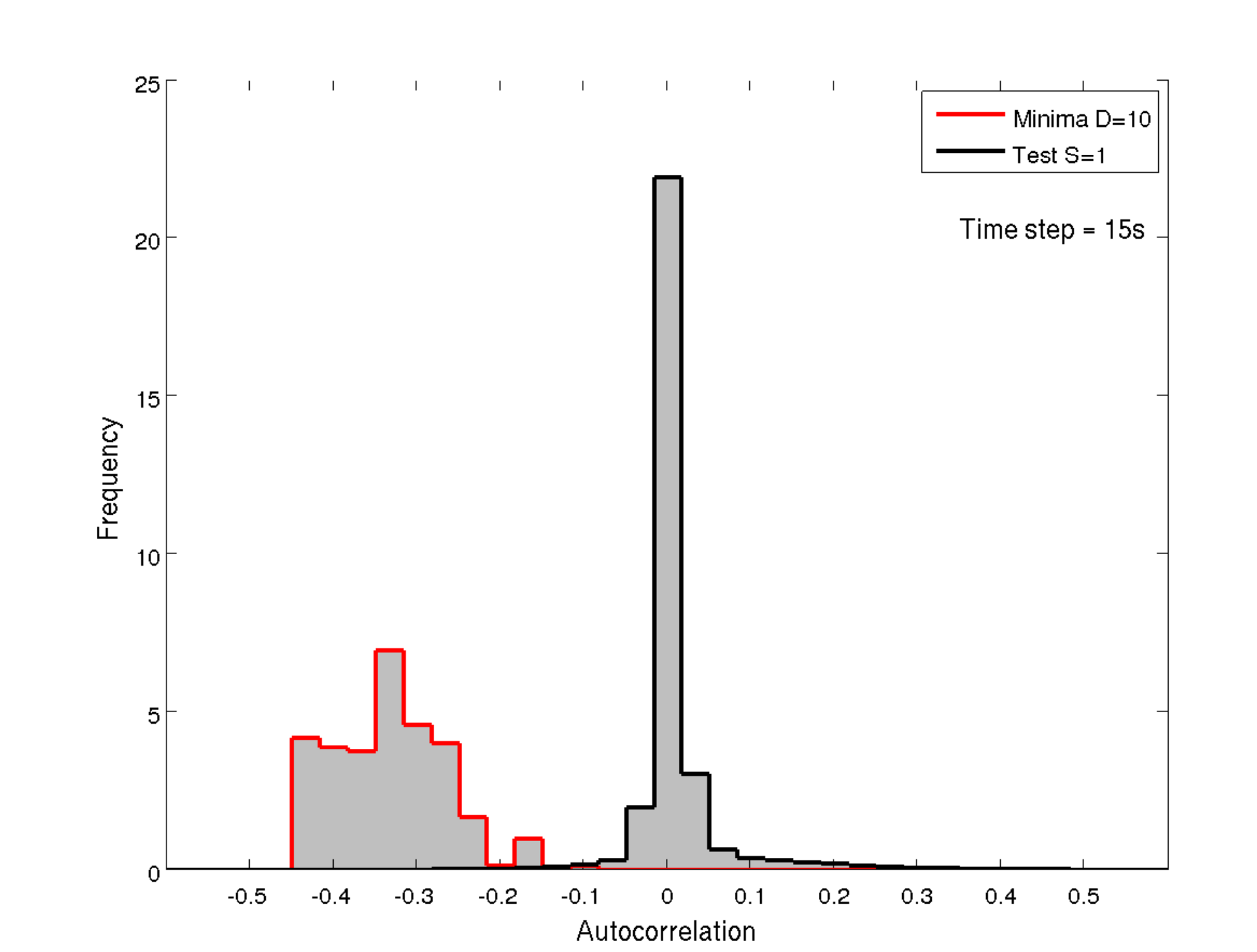}
\caption{Empirical distribution of minimal values obtained during $D=10$ consecutive days (red) and distribution of single entries the corresponding matrix $\mathbf{\hat{C}}$ (black). The depicted case corresponds to a time scale of $15s$; the remaining ones keep qualitatively the same look.}
\label{fig:crossauto}
\end{figure}

\section{Results}

The main results are reported in Fig.~\ref{fig:res1} and \ref{fig:res2}. Each distinct picture contains the data for a distinct time scale. The obtained autocorrelation values are reported as histograms. The red stands for the values during the minimization periods, blue and black for, respectively, the test with $S=1$ and $S=5$. The vertical lines indicate the corresponding null hypothesis. Fig.~\ref{fig:meanvalues} plots the mean values of these distributions as functions of the time scale.

The minimal values are far below those obtained from the synthetic independent time series (see Fig.~\ref{fig:white}). There is not much deviation among the different time scales, if only in the variance of the distributions. The mean value does not experience large deviations. The only time scale where we observe a smaller effect is $5s$. However, the desired property appears more strikingly for time scales of 10 and $15s$.

The autocorrelations computed during the test periods deviate from the minimized values, with differences that become stronger for higher time scales. In this case too, the time scales of 10 and $15s$ exhibit the most important signal. Fig.~\ref{eq:significative} shows the fraction of values that violate the null hypothesis, i.e. the values that are not compatible with independence as defined in Sec.~\ref{sec:autocorrelation}. For the longer test and for the time scale equal to $15s$, this fraction is about 0.9.

In the same fashion of Fig.~\ref{fig:meanvalues}, Fig.~\ref{fig:hurst_results} reports the mean values of the estimated Hurst exponent. In order to provide a comparison, Fig.~\ref{fig:hurst_rand} presents the distribution of the Hurst exponents estimated by randomly extracting 10 consecutive days of DJIA data. The mean value of this distribution is slightly smaller than 0.5 and it is equal to 0.494. In general, the values of $H$ we obtain in Fig.~\ref{fig:hurst_results} are significantly different from 0.5, with the stronger effects for time scales of about $20~35s$.

A study of the weight dynamics is beyond the purpose of this paper. However, Fig.~\ref{fig:weights} reports the empirical weight distribution with an Euclidean normalization. The pick around zero shows the optimal basket is often not formed by all the considered stocks, but the remaining part suggests we are effectively observing collective dynamics rather than just handling a small subset of the assets. 

Finally, Fig.~\ref{fig:crossauto} reports the distribution of the entries of  $\mathbf{\hat{C}}$, i.e. the autocorrelations and the lead-lag cross-correlations, together with the found minima, showing that the ``distance'' between those two samples is strong an suggesting one more time the collective nature of the measured effect. In both Fig.~\ref{fig:weights} and Fig.~\ref{fig:crossauto} the time scale is $15s$ and the qualitative result does not change in the other cases.

\section{Conclusion}
A clear answer to the main research question of this work has been given. In fact, at short time scales it is possible to build a basket that does not fit into the efficient market paradigm. Since we have analyzed mid-price time series, it is not possible, at this stage, to ascertain whether this effect is strong enough to be exploited for trading purposes. Preliminary results in this direction suggest that this practice is not trivial but yet possible, i.e. the fluctuations we observe often fall within the bid-ask spread, but this is not always the case. Another approach, perhaps using non-parametric optimization methods such as a stochastic optimization algorithm, is likely to provide stronger results. Moreover, we have just shown the values for the optimized baskets, but it is likely that penalizing the optimal values with some other suitable criterion would lead to a more favorable situation in terms of trading applications - for example, one can try to reproduce this study, but this time considering baskets which minimize the autocorrelation only among those with a minimal given value of volatility. Moreover, the value of $D$ for which we have shown the results can be tuned, and so can the time scales. Approaching the problem without imposing a time step would probably lead to a much more complete view of the effect. 

Finally, we must also point out the fact that we considered all the 30 DJIA stocks, without developing any smart way of picking them. Since we already have clear results using the full index, we are keen to think that a smarter choice of the companies would lead to stronger effects. 

\section*{Acknowledgement}
F.A. would like to acknowledge that the idea of optimizing a basket of stocks with respect to statistical properties arose from a collaboration with Laurent Jaillet while they were colleagues at CAI Cheuvreux.

\bibliographystyle{plain}
\bibliography{paper}

\end{document}